\documentclass[11pt,a4paper]{article}
\pdfoutput=1
\usepackage{jcappub}
\usepackage{bm}
\usepackage{comment}
\usepackage{amsmath}
\usepackage{amssymb}
\usepackage{booktabs}
\usepackage{graphicx}
\usepackage{float}
\usepackage{color}

\begin{document}
\begin{sloppypar}
\title{Sky localization of space-based detectors with time-delay interferometry}

\author[a,b]{Tong Jiang,}
\author[b,c,1]{Yungui Gong\note{Corresponding author.},}
\author[b]{Xuchen Lu}

\affiliation[a]{College of Physics and Technology, Kunming University, 2 Puxin Rd, Kunming, Yunnan 650214, China}
\affiliation[b]{School of Physics, Huazhong University of Science and Technology, 1037 LuoYu Rd, 
Wuhan, Hubei 430074, China}
\affiliation[c]{Institute of Fundamental Physics and Quantum Technology, Department of Physics, School of Physical Science and Technology, Ningbo University, 818 Fenghua Rd, Ningbo, Zhejiang 315211, China}

\emailAdd{jiangtong@hust.edu.cn}
\emailAdd{gongyungui@nbu.edu.cn}
\emailAdd{Luxc@hust.edu.cn}


\abstract{The accurate sky localization of gravitational wave (GW) sources is an important scientific goal for space-based GW detectors.
The main differences between future space-based GW detectors, such as Laser Interferometer Space Antenna (LISA), Taiji, and TianQin, include the time-changing orientation of the detector plane, 
the arm length, the orbital period of the spacecraft and the noise curve.
Because of the effects of gravity on three spacecraft, 
it is impossible to maintain the equality of the arm length,
so the time-delay interferometry (TDI) method is needed to cancel out the laser frequency noise for space-based GW detectors.
Extending previous work based on equal-arm Michelson interferometer, 
we explore the impacts of different first-generation TDI combinations and detector's constellations on the sky localization for monochromatic sources.
We find that the sky localization power is almost unaffected by the inclusion of the TDI Michelson $(X, Y, Z)$ combination in the analysis.
We also find that the variation in the sky localization power for different TDI combinations is entirely driven by the variation in the sensitivities of these combinations.
For the six particular TDI combinations studied,
the Michelson $(X, Y, Z)$ combination is the best for source localization.}

\arxivnumber{2301.05923}

\maketitle

\section{Introduction}
The first detection of gravitational waves (GWs) \cite{LIGOScientific:2016aoc} by advanced Laser Interferometer Gravitational-Wave Observatory (LIGO) Scientific Collaboration \cite{Harry:2010zz, LIGOScientific:2014pky} and Virgo Collaboration \cite{VIRGO:2014yos} opened a new window to test Einstein's general relativity \cite{LIGOScientific:2016lio, LIGOScientific:2018dkp, LIGOScientific:2019fpa, LIGOScientific:2020tif},
study the population of stellar-mass black holes,
measure the merger rates of binary black holes in the Universe, 
probe the evolution history of the Universe
and understand the nature of the gravity \cite{LIGOScientific:2016dsl,LIGOScientific:2016kwr,LIGOScientific:2016hpm,LIGOScientific:2021aug,Seto:2001qf,Kyutoku:2016zxn,eLISA:2013xep,Klein:2015hvg,Sesana:2008ur,Ricarte:2018mzn,Li:2022fno,Mangiagli:2022niy}.
However, the ground-based GW observatories are only sensitive to GWs in the high frequency bands $10-10^3$ Hz due to seismic noise and gravity gradient noise.
The proposed space-based GW detectors such as Laser Interferometer Space Antenna (LISA) \cite{Danzmann:1997hm, LISA:2017pwj}, Taiji \cite{Hu:2017mde}, and TianQin \cite{TianQin:2015yph} will detect GWs in the low frequency regimes $10^{-4}-10^{-1}$ Hz where a wealth of astrophysical sources resides.
Accurate sky localization is an important scientific goal for GW observations \cite{Grover:2013sha}.
The accurate information about the source
localization is essential for the
follow-up observations of electromagnetic counterparts and the statistical
identification of the host galaxy when no counterpart is
present \cite{Cutler:1997ta,Grover:2013sha,Berry:2014jja},
so that GWs can be used as standard sirens \cite{Schutz:1986gp, Holz:2005df} to probe the evolution of the Universe and study the problem of the Hubble tension \cite{Chen:2017rfc,Hotokezaka:2018dfi,Vitale:2018wlg,LIGOScientific:2019zcs,Zhu:2021bpp,Riess:2019cxk}.
Moreover, accurate knowledge about the GW source position
may provide important information about the environments where the binary source resides.
However, the errors in determining the source position are strongly correlated with the other parameters of the binary source \cite{Cutler:1997ta,Blaut:2011zz,Zhang:2023ceh},
so accurate source localization depends on the GW detector and the type of source.
In contrast to the coalescing signals detectable by ground-based detectors,
space-based GW detectors can measure monochromatic GWs from short-period binary stars for months to years,
and the translatory motion of detector's center around the Sun imposes on the signal a periodic Doppler shift, 
the Doppler modulation on the amplitude and phase of GWs carries the position information of the sources.
Therefore, a single space-based GW detector is able to locate the source position.

There are two different constellations for space-based GW detectors.
For LISA/Taiji, the spacecraft is in the heliocentric orbit behind/ahead of the Earth by about $20^\circ$,
the inclination angle between the plane of constellations and ecliptic plane is $60^\circ$, and they keep the geometry of an almost equilateral triangle with the average arm length $L=2.5\times10^9$ m$/3\times10^9$ m.
For TianQin, the spacecraft is in the geocentric orbit around the Earth and rotates around the Sun,
the arm length is $L=1.73\times10^8$ m, and
the normal vector of the detector plane points to the source RX J0806.3+1527 at ($\theta_{tq}=-4.7^{\circ}$, $\phi_{tq}=120.5^{\circ}$).
The main differences between different constellations of space-based GW detectors are the time-changing orientation of the detector plane,
the arm length, the orbital period of the spacecraft and the noise curve \cite{Zhang:2020hyx}.
It is necessary to discuss the effects of
these factors on the accuracy of source localization.

As discussed in \cite{Zhang:2020hyx,Zhang:2020drf,Gong:2021gvw} for monochromatic sources, 
the source-position-dependent modulation on the signal amplitude due to the rotation of the detector plane with a period of one year which dominates over the Doppler modulation at frequencies below 1mHz, not only helps LISA and Taiji get better accuracy in the sky localization of GW source but also increases the sky coverage at frequencies below 1mHz \cite{Cutler:1997ta,Blaut:2011zz,Zhang:2020hyx,Zhang:2020drf,Gong:2021gvw,Peterseim:1996cw}.
The ability in the sky localization of TianQin is better than LISA and Taiji at frequencies above 30 mHz
and TianQin has blind spots for sources from the directions with $\phi_s$ around $30^\circ$ or $-150^\circ$.
At higher frequencies when the wavelength of GWs is smaller
than the detector's arm length, 
the frequency-dependent transfer function
deteriorates the signal-to-noise ratio (SNR) registered in the detector.
Compared to the individual detectors, 
the network of LISA and TianQin has a better ability in sky localization for sources with frequencies in the range of 1-100 mHz and offers larger sky coverage for angular resolutions.

The motion of spacecraft in space makes space-based GW detectors impossible to maintain the exact equality between the arm lengths.
Because of the unequal arm lengths,
the laser frequency noise cannot be cancelled out completely.
To reduce the laser frequency noise, properly chosen time shifted and linearly combined data streams were proposed to synthesize virtual equal arm interferometric measurements \cite{Tinto:1994kg,Tinto:1999yr,Armstrong_1999}.
This technique is known as time-delay interferometry (TDI) \cite{Tinto:1999yr,Armstrong_1999,Estabrook:2000ef,Tinto:2003vj,Vallisneri:2004bn}.
The first-generation TDI combinations can cancel out the laser frequency noise in a static unequal-arm configuration,
and the second-generation TDI can cancel out the laser frequency noise in a rotating and flexing configuration with arm lengths varying linearly in time
\cite{Shaddock:2003dj,Tinto:2003vj,Cornish:2003tz}.
The first-generation TDI configurations were applied for space-based GW detectors such as LISA, Taiji and TianQin in \cite{Estabrook:2000ef,Tinto:2001ii, Tinto:2001ui,Hogan:2001jn,Armstrong:2001uh,Prince:2002hp,Tinto:2002de,Shaddock:2003dj,Tinto:2003uk,Nayak:2003na,Tinto:2004nz,Romano:2006rj,Zhang:2020khm},
and the applications of second-generation TDI configurations were discussed in \cite{Tinto:2003vj,Cornish:2003tz,Vallisneri:2004bn,Krolak:2004xp,Vallisneri:2005ji,Wang:2017aqq,Wang:2020fwa,Wang:2020pkk,RajeshNayak:2004jzp,Nayak:2005un}.
For more discussion on TDI algorithm and its application to space-based GW detectors, please see \cite{Tinto:2004wu,Tinto:2014lxa,Tinto:2020fcc,Muratore:2020mdf} and references therein.
Since the noises in different TDI combinations are different, 
we expect that the ability of sky localization with different TDI combinations will be different. 
Thus, we extend previous work to study the effect of different TDI combinations on the sky localization. 

The paper is organized as follows.
In Sec. \ref{meth}, we provide a brief overview of TDI response and Fisher information matrix (FIM) method.
In Sec. \ref{result}, we consider five fiducial detectors and six particular TDI combinations to discuss the effects of different TDI combinations and constellation factors on the sky localization for monochromatic GWs,
and compare the results of angular resolutions between LISA, Taiji, TianQin, and their network.
Finally, we present the discussion in Sec. \ref{discussion}.

\section{Methodology}\label{meth}
In this section, we provide a brief overview on TDI responses of space-based GW detectors to monochromatic GWs and parameter estimations with the method of FIM.

\subsection{The GW signal}
\label{wave}
For GWs propagating along the direction $\hat{\Omega}(\theta,\phi)$, we can define two perpendicular unit vectors $\hat{p}$ and $\hat{q}$ that satisfy the orthogonal relation $\hat{\Omega}=\hat{p}\times \hat{q}$.
To account for the rotational degree of freedom around $\hat{\Omega}$, we introduce the polarization angle $\psi$ to form two new orthonormal vectors
\begin{equation}
    \hat{r}=\cos(\psi) \hat{p}+\sin(\psi)\hat{q},\qquad\hat{s}=-\sin(\psi) \hat{p}+\cos(\psi)\hat{q}.
\end{equation}
With these two orthonormal bases, the plus $(+)$ and cross $(\times)$ polarization tensors are
\begin{equation}
    e^+_{ij}=\hat{r}_i\hat{r}_j-\hat{s}_i\hat{s}_j, \qquad e^\times_{ij}=\hat{r}_i\hat{s}_j+\hat{r}_j\hat{s}_i,
\end{equation}
and the GW signal can be expressed as
\begin{equation}
h_{ij}(t)=\sum_{A=+,\times} e_{ij}^A h_A(t).   
\end{equation}
For monochromatic GWs with the frequency $f_0$ emitted from sources such as compact binaries containing stellar or intermediate-mass black holes, white dwarfs or neutron stars, in the lowest order of quadrupole approximation the tensor modes are
\begin{equation}
\label{monowave}
\begin{split}
   &h_+(t) = \mathcal{A}[1+ \cos^2(\iota)]e^{(2\pi i f_0 t + i \phi_0)},\\
   &h_\times(t) = 2i \mathcal{A} \cos^2(\iota) e^{(2\pi i f_0 t + i \phi_0)},
\end{split}
\end{equation}
where the amplitude $\mathcal{A}=2m_1 m_2(\pi f_0)^{2/3}/[(m_1+m_2)^{1/3} d_L]$, $m_1$ and $m_2$ are masses of the binary components, $d_L$ is the luminosity distance between the source and the observer, $\phi_0$ is the initial phase of the orbital plane and $\iota$ is the inclination angle of the detector plane.
The signal registered in the detector $\alpha$ is
\begin{equation}
\label{signal}
    z_{\alpha}(t)=\sum_A F^A_{\alpha}(f,\theta,\phi,\psi) h_A(t) e^{i\phi_D(t)} + n_\alpha(t),
\end{equation}
where $n_\alpha(t)$ is the detector noise, $\phi_D(t)$ is the Doppler phase, and the $F^A_{\alpha}(f,\theta,\phi,\psi)$ is the response function in the detector $\alpha$ \cite{Estabrook:2000ef}. 
For GW signals with the frequency $f_0$, the Doppler phase is 
\begin{equation}
\label{dphase}
    \phi_{D}(t)=2 \pi f_0 R_s \sin \theta \cos \left(\frac{2 \pi t}{P}-\phi-\phi_{\alpha}\right),
\end{equation}
where $R_s=1$ AU is the distance between the sun and the earth, $\theta$ and $\phi$ are the angular position of the GW source in the ecliptic coordinate, $P$ is the period of rotation for the earth around the sun, and $\phi_{\alpha}$ is the detector's ecliptic longitude at $t=0$, so the source parameters are  $\bm{\Lambda} = \left(\theta,\ \phi,\ \mathcal{A},\ \iota,\ \psi,\ \phi_0\right)$.

\subsection{The TDI configurations and fiducial detectors}\label{instrument}
\begin{figure}[htbp]
    \centering
    \includegraphics[width=0.8\columnwidth]{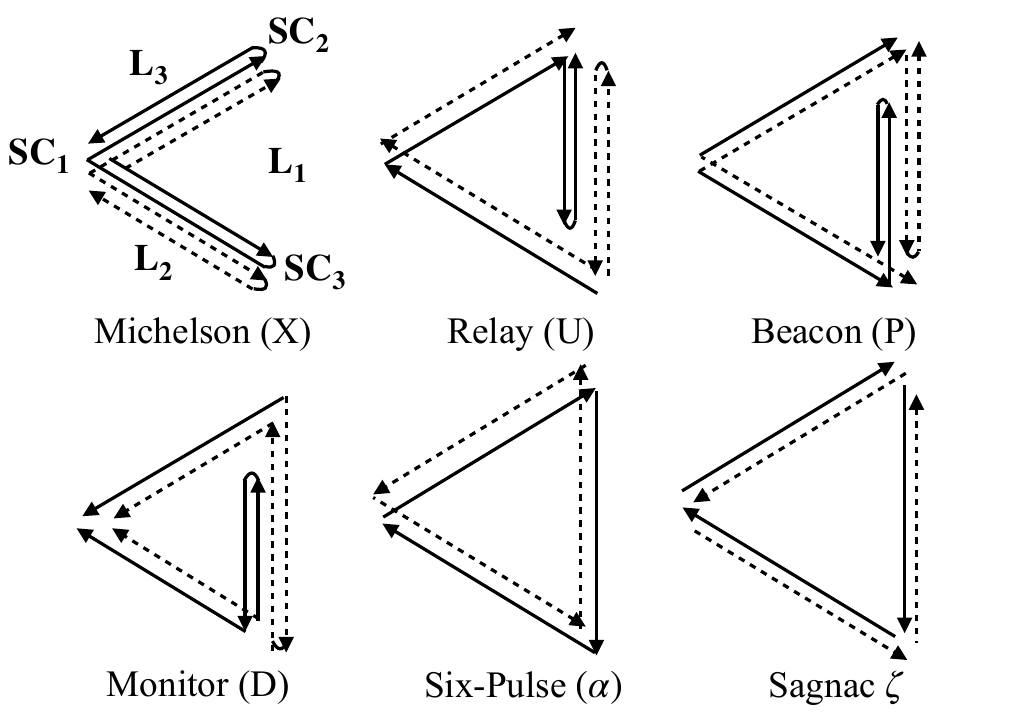}
    \caption{The configurations of six first-generation TDI combinations \cite{Estabrook:2000ef,Vallisneri:2005ji}.}
    \label{schematic}
\end{figure}

For space-based GW detectors, it is impossible to maintain the exact equality of the arm lengths due to the motion of spacecraft (SC), 
so laser frequency noise cannot be effectively removed when two beams are recombined directly
at the photo-detector.
Fortunately, we can form, from the multiple readouts of space-based GW detectors, 
observables that are insensitive to laser frequency fluctuations and optical bench motions by properly choosing time shifted and linearly combined data streams.
This technique is known as TDI.
Although there are numerous TDI combinations \cite{Muratore:2020mdf}
which can be derived from the four generators $(\alpha, \beta, \gamma,\zeta)$ \cite{Prince:2002hp},
we focus on six particular TDI combinations \cite{Tinto:2004wu,Tinto:2014lxa,Tinto:2020fcc} in this paper.
As shown in figure \ref{schematic}, these six combinations for the first-generation TDI are
Sagnac $\zeta$, six-pulse $(\alpha, \beta, \gamma)$,  Michelson $(X, Y, Z)$, 
Relay $(U, V, W)$, Beacon $(P, Q, R)$, and monitor $(E, F, G)$.
In figure \ref{schematic}, the spacecraft is labeled as 1, 2, 3, 
the optical paths between two SCs are denoted by $L_a$ which are assumed to be constant for the discussion of the first-generation TDI combinations, 
the index $a$ corresponds to the opposite $\text{SC}_a$ and $a=1,2,3$.
The three channels of each configuration are obtained by cyclical permutation of the spacecraft indices.

\begin{figure}[htbp]
    \centering
    \includegraphics[width=0.6\columnwidth]{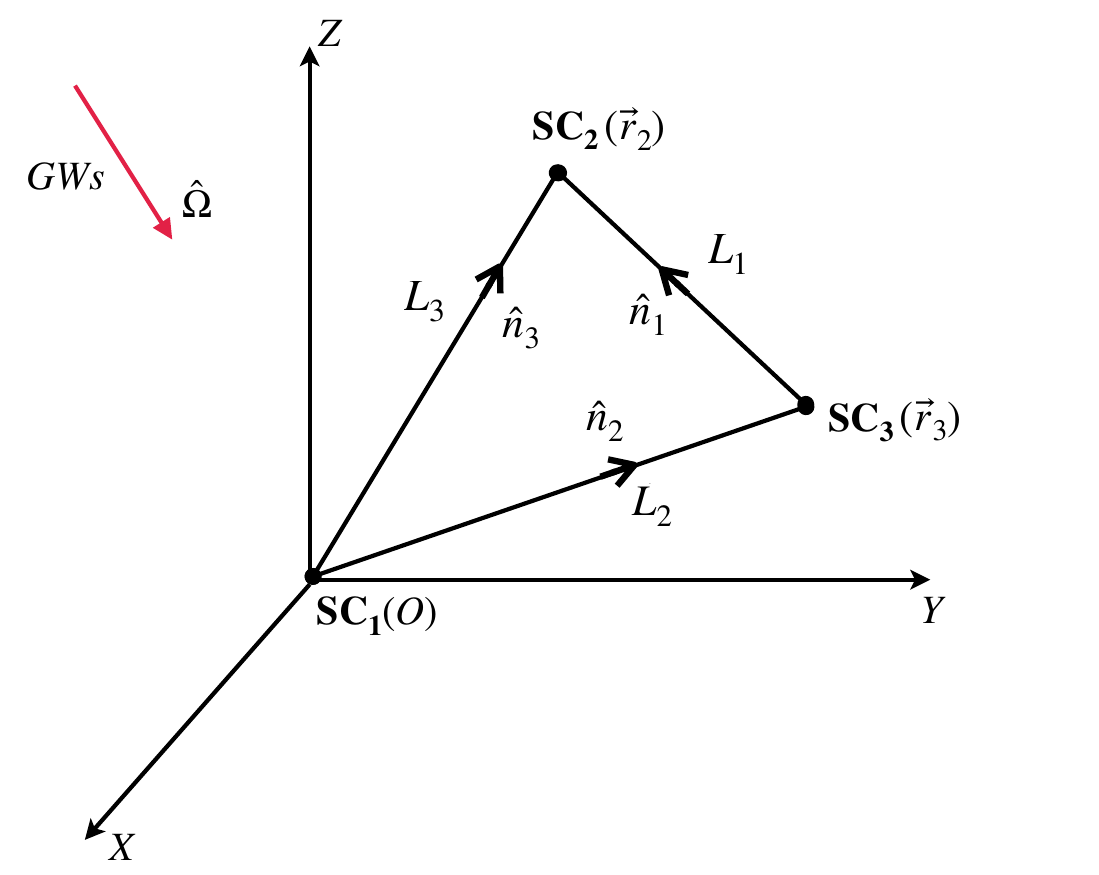}
    \caption{The one-way Doppler tracking.
    We put $\text{SC}_1$ at the origin of the coordinate, 
    the other two spacecraft $\text{SC}_2$ and $\text{SC}_3$ are located at $\vec{r}_2$ and $\vec{r}_3$.
    GWs propagate along the direction $\hat{\Omega}$, the optical paths between two SC pairs are denoted by $L_a$ and the unit vectors along the path are $\hat{n}_a$, with the index $a$ corresponding to the opposite SC.}
    \label{schematic2}
\end{figure}

For the response of the one-way transmission, as shown in figure \ref{schematic2}, between $\text{SC}_3$ and $\text{SC}_2$ with the distance $L_1$ along the unit direction $\hat{n}_1$ is \cite{Larson:2002xr}
\begin{equation}
\label{resp1}
\begin{split}
	\delta L_1&=\sum_{A,i,j}\hat{n}^i_1\hat{n}^j_1 e^A_{ij}\frac{\sin[\omega L_1(1-\hat{n}_1\cdot\hat{\Omega})/2]}{\omega(1-\hat{n}_1\cdot\hat{\Omega})}e^{-i\omega L_1(1-\hat{n}_1\cdot\hat{\Omega}+2\Omega\cdot\vec{r}_2/L_1)/2}h^A(f)\\
&=\sum_{A,i,j}L_1 \hat{n}^i_1\hat{n}^j_1 e^A_{ij}T(\omega,\hat{n}_1\cdot\hat{\Omega})h^A(f),
\end{split}
\end{equation}
where $\omega=2\pi f$ is the angular frequency of GWs, $\hat{\Omega}$ is the propagating direction of GWs.
We take the speed of light $c=1$.
The transfer function $T(\omega,\hat{n}_1\cdot\hat{\Omega})$ is \cite{Cornish:2001qi}
\begin{equation}
\label{transf1}
\begin{split}
T(\omega,\hat{n}_1\cdot\hat{\Omega})&=\frac{\sin[\omega L_1(1-\hat{n}_1\cdot\hat{\Omega})/2]}{\omega L_1(1-\hat{n}_1\cdot\hat{\Omega})}e^{-i\omega L_1(1-\hat{n}_1\cdot\hat{\Omega}+2\Omega\cdot \vec{r}_2/L_1)/2}\\
&=\frac{1}{2}\text{sinc}\left[\omega L_1(1-\hat{n}_1\cdot\hat{\Omega})/2\right]e^{-i\omega L_1(1-\hat{n}_1\cdot\hat{\Omega}+2\Omega\cdot \vec{r}_2/L_1)/2},
\end{split}
\end{equation}
where $\text{sinc}(x)=\sin{x}/x$ and $\vec{r}_2$ is the location of $\text{SC}_2$. In the long-wavelength limit, $\omega L_1\rightarrow 0$,
$\text{sinc}\left[\omega L_1(1-\hat{n}_1\cdot\hat{\Omega})/2\right]\approx 1$ which is independent of the GW frequency.

Following \cite{Armstrong_1999},
we denote $y_{ab}$ as the relative frequency
fluctuations time series measured from reception at $\text{SC}_b$ with transmission from $\text{SC}_d$ ($d\ne a$ and $d\ne b$) along $L_a$,
for example, $y_{31}$ is the relative frequency fluctuations time series measured at spacecraft 1 with transmission from spacecraft 2 along $L_3$;
we also use the notation $y_{ab,c}$ for the delayed data stream, 
for example, $y_{31,2}=y_{31}(t-L_2)$ and $y_{31,32}=y_{31}(t-L_3-L_2)$.
By using Eqs. (\ref{resp1}) and (\ref{transf1}), the six TDI signal responses of GW are 
\begin{equation}
\begin{split}
    y_{ab}\left(T_D\right)&=-\frac{\delta L_a}{L_a}=-\sum_{A, i, j} \hat{n}_a^i \hat{n}_a^j e_{i j}^A T_{a b}\left(T_D\right) h^A(f),\\
    T_{12}\left(T_D\right)&=\frac{1}{2} \operatorname{sinc}\left[\frac{1}{2} u_1\left(1-\mu_1\right)\right] e^{-i u_1\left(1-\mu_1\right) / 2-i\left(\mu_3 u_3+T_D\right)},\\
    T_{23}\left(T_D\right)&=\frac{1}{2} \operatorname{sinc}\left[\frac{1}{2} u_2\left(1-\mu_2\right)\right] e^{-i u_2\left(1-\mu_2\right) / 2-i\left(\mu_2 u_2+T_D\right)},\\
    T_{32}\left(T_D\right)&=\frac{1}{2} \operatorname{sinc}\left[\frac{1}{2} u_3\left(1-\mu_3\right)\right] e^{-i u_3\left(1-\mu_3\right) / 2-i\left(\mu_3 u_3+T_D\right)},\\
    T_{31}\left(T_D\right)&=\frac{1}{2} \operatorname{sinc}\left[\frac{1}{2} u_3\left(1+\mu_3\right)\right] e^{-i u_3\left(1+\mu_3\right) / 2-i T_D},\\
    T_{13}\left(T_D\right)&=\frac{1}{2} \operatorname{sinc}\left[\frac{1}{2} u_1\left(1+\mu_1\right)\right] e^{-i u_1\left(1+\mu_1\right) / 2-i\left(\mu_2 u_2+T_D\right)},\\
    T_{21}\left(T_D\right)&=\frac{1}{2} \operatorname{sinc}\left[\frac{1}{2} u_2\left(1+\mu_2\right)\right] e^{-i u_2\left(1+\mu_2\right) / 2-i T_D},
\end{split}
\end{equation}
where $u_a=(2\pi f) L_a$, $\mu_a=\hat{n}_a \cdot \hat{\Omega}$, $T_D$ is the corresponding time delay, for example, $T_D=u_1+u_2$ for $y_{ab,12}$. 
The response function $F^A(f,\theta,\phi,\psi)= \hat{n}_a^i \hat{n}_a^j e_{i j}^A T_{a b}\left(T_D\right)$.

Due to the cancellation of frequency noises in the stationary unequal-arm interferometry by the first-generation TDI combinations, 
the dominant detector noises are the acceleration noise and the single-link optical metrology noise.
The acceleration noises for LISA \cite{Robson:2018ifk} and Taiji \cite{Wang:2020fwa} are 
\begin{equation}
\begin{split}
    &S^{a}_\text{LISA} = S^{a}_\text{TJ}=(3\times 10^{-15}\  \text{m s}^{-2})^2 \bigg[1+\big(\frac{0.4 \ \text{mHz}}{f}\big)^2\bigg]\bigg[1+\big(\frac{f}{8 \ \text{mHz}}\big)^4\bigg]\ \text{Hz}^{-1},
\end{split}
\end{equation}
the single-link optical metrology noises for LISA and Taiji are
\begin{equation}
\begin{split}
    &S^{x}_\text{LISA} = (1.5\times 10^{-11}\  \text{m})^2 \bigg[1+\big(\frac{2 \ \text{mHz}}{f}\big)^4\bigg]\ \text{Hz}^{-1},\\
    &S^{x}_\text{TJ} = (8\times 10^{-12}\  \text{m})^2 \bigg[1+\big(\frac{2 \ \text{mHz}}{f}\big)^4\bigg]\ \text{Hz}^{-1},
\end{split}
\end{equation}
respectively. 
The acceleration noise and the single-link optical metrology noise
for TianQin are \cite{TianQin:2015yph}
\begin{equation}
    \begin{split}
       &S^a_\text{TQ}=10^{-30}\text{m}^2\ \text{s}^{-4}/\text{Hz},\\
       &S^x_\text{TQ}=10^{-24}\text{m}^2/\text{Hz}.
    \end{split}
\end{equation}
For an equal-arm detector with the arm length $L$, 
the fractional frequency fluctuation spectra for the shot and proof mass noises are related to the single-link optical metrology and acceleration noises as \cite{Armstrong_1999,Estabrook:2000ef}
\begin{equation}
\label{tdinoise1}
\begin{split}
&S_y^{\rm shot}=S^{x}/L^2,\\
&S_y^{\rm proof\ mass}=S^{a}/(L^2(2\pi f)^4),
\end{split}
\end{equation}
where the subscript $y$ represents the individual laser link $y_{ab}$. 
For LISA, $L=2.5\times10^9$ m; for Taiji, $L=3\times10^9$ m; and for TianQin, $L=1.78\times10^8$ m.
For the space-based GW detectors, the three laser arms are treated as two Michelson interferometers during the detection process, with one laser arm being shared by both.
This introduces correlation issues in signal analysis, causing the noise to become partially correlated \cite{Prince:2002hp,Vallisneri:2007xa,Tinto:2014lxa,Muratore:2022nbh}.
However, the impact of noise correlation is the higher-order effect in measurement, even though the difference of the power spectral density of noise between full-correlated and none correlated noises is similar in the case of Michelson $(X, Y, Z)$ combination \cite{Nam:2022rqg}.
Therefore, in this paper, we simplify the calculation by treating the noise performance as the uncorrelated single-link contribution from the optical measurement system and test-mass acceleration.

The noises in different TDI combinations are \cite{Estabrook:2000ef}
\begin{equation}
\label{pnalpha}
P_n^\alpha=[8\sin^2(3\pi f L)+16\sin^2(\pi fL)]S_y^{\rm proof\ mass}+6 S_y^{\rm shot},
\end{equation}

\begin{equation}
\label{pnzeta}
P_n^\zeta=24\sin^2(\pi fL) S_y^{\rm proof\ mass}+6S_y^{\rm shot},
\end{equation}

\begin{equation}
\label{pnx}
P_n^X=[8\sin^2(4\pi f L)+32\sin^2(2\pi fL)]S_y^{\rm proof\ mass}+16\sin^2(2\pi fL)S_y^{\rm shot},
\end{equation}

\begin{equation}
\label{pnp}
P_n^P=[8\sin^2(2\pi f L)+32\sin^2(\pi fL)]S_y^{\rm proof\ mass}+[8\sin^2(2\pi fL)+8\sin^2(\pi fL)]S_y^{\rm shot},
\end{equation}

\begin{equation}
\label{pne}
P_n^E=[32\sin^2(\pi f L)+8\sin^2(2\pi fL)]S_y^{\rm proof\ mass}+[8\sin^2(\pi fL)+8\sin^2(2\pi fL)]S_y^{\rm shot},
\end{equation}

\begin{equation}
\label{pnu}
\begin{split}
P_n^U=&[16\sin^2(\pi f L)+8\sin^2(2\pi fL)+16\sin^2(3\pi fL)]S_y^{\rm proof\ mass}\\
&+[4\sin^2(\pi fL)+8\sin^2(2\pi fL)+4\sin^2(3\pi fL)]S_y^{\rm shot}.
\end{split}
\end{equation}

Since GWs come from all directions, 
we use the average response function $R(f)$
by taking the average of the response function $F^A$ over all source directions and the polarization angle \cite{Zhang:2019oet}.
The sensitivity curve is $S_n(f)=P_n(f)/R(f)$.

\begin{figure}[htbp]
    \centering
    \includegraphics[width=0.98\columnwidth]{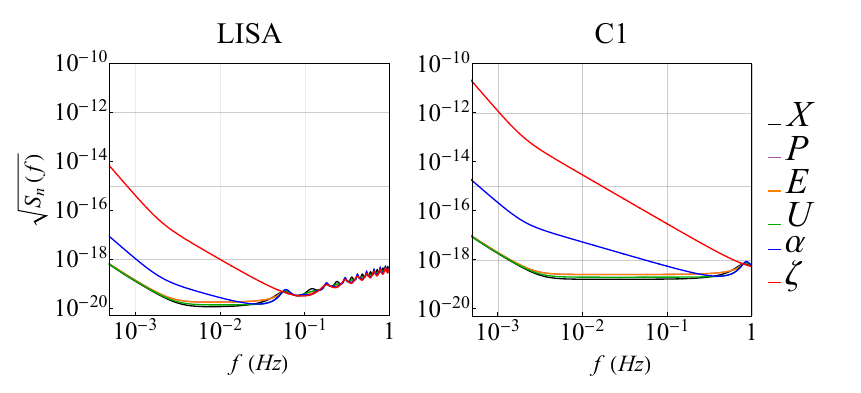}
    \caption{The sensitivity curves $\sqrt{S_n(f)}$ of LISA and C1 for the six TDI combinations for the tensor modes.}
    \label{sensitivity}
\end{figure}

For different constellations of space-based GW detectors, three dominant design factors affect the accuracy of source localization:
the rotation, the arm length, and the rotation period of the spacecraft \cite{Zhang:2020hyx}.
To examine the effects of these factors individually, 
we use the control variable method and construct five fiducial GW detectors as shown in table \ref{fdetector} \cite{Zhang:2020hyx}. 
The LISA-like fiducial detectors R and R1 are in the heliocentric orbit with the orbital period of one year and the normal vector of the detector plane rotates around the normal vector of the ecliptic plane with a period of one year. 
The TianQin-like fiducial detectors C, C1 and C2 are in the geocentric orbit and the normal vector of the detector plane points to the calibration source RX J0806.3+1527.
The orbital period of the fiducial detectors C and C1 is one year while the orbital period of the fiducial detector C2 is 3.65 days.
The detailed orbit equations for these fiducial detectors were given in Ref. \cite{Zhang:2020hyx}.
The fiducial detector R (R1) and C (C1) are constructed to study the effect of rotation,
the fiducial detector R (C) and R1 (C1) are constructed to study the effect of the arm length, and
the fiducial detector C1 and C2 are constructed to study the effect of the rotation period of the spacecraft.
For all fiducial detectors, we assume that the acceleration noise and the single-link optical metrology noise are the same as LISA.
In figure \ref{sensitivity}, we plot $\sqrt{S_n(f)}$ of LISA and the fiducial detector C1 for the tensor modes for the six TDI combinations.
From figure \ref{sensitivity}, we see that the Sagnac combination $\zeta$ suppresses the GW signals because its response to GWs becomes of higher order \cite{Armstrong_1999}, especially below $0.1$ Hz, 
so we expect that the sky localization will be very bad with the $\zeta$ combination.

\begin{table}\centering
\begin{tabular}{ |p{3cm}||c| c| c| c| c| c|}
    \hline
    Detector  & $\text{R}$ & $\text{C}$  &$\text{R1}$  & $\text{C1}$ &$\text{C2}$\\
    \hline
    Arm Length (m) & $3.7\times10^9$ & $3.7\times10^9$  &$1.73\times10^8$  & $1.73\times10^8$ &$1.73\times10^8$\\
    \hline
    Period  & $1\ \text{year}$ & $1\ \text{year}$  &$1\ \text{year}$  & $1\ \text{year}$ &$3.65\ \text{days}$\\
    \hline
\end{tabular}
\caption{The parameters of the fiducial detectors. The LISA-like fiducial detectors R and R1 are in the heliocentric orbit, 
the TianQin-like fiducial detectors C, C1 and C2 are in the geocentric orbit and their normal vectors of the detector planes point to the calibration source RX J0806.3+1527.
}
\label{fdetector}
\end{table}

\subsection{FIM method}
\label{analysis}
Given two frequency-domain signals $h_1(f)$ and $h_2(f)$, the inner product $\langle h_1|h_2\rangle$ is defined as
\begin{equation}\label{overlap}
    \langle h_1|h_2\rangle=2\int_0^{+\infty } \frac{h_1(f) h_2^*(f)+ h_2(f) h_1^*(f)}{ P_n(f)}\,df,
\end{equation}
where  $h^*$ denotes the complex conjugate of $h$.
The SNR of GW signal $z_\alpha$ is
\begin{equation}
\label{snr}
\begin{split}
\rho^2=& \left< z_\alpha | z_\alpha \right>\\
=&4\  \int_{0}^{\infty}\frac{ z_\alpha(f) z_\alpha^*(f)}{P_{n}^\alpha(f)}df.
\end{split}
\end{equation}
In this paper, we choose the threshold of detecting a signal as $\rho\ge 10$.
To estimate the uncertainty of source parameters, we apply the FIM method.
The FIM is defined as
\begin{equation}
\label{tgamma}
\begin{split}
\Gamma_{ij}=& \left<\frac{\partial z_\alpha}{\partial \bm{\Lambda}_i}\left|\frac{\partial z_\alpha^*}{\partial \bm{\Lambda}_j}\right.\right>\\
=&4\ \text{Re} \int_{0}^{\infty}\frac{\partial_i z_\alpha(f)\partial_j z_\alpha^*(f)}{P_{n}^\alpha(f)}df\\
=&\frac{2}{P_{n}^\alpha(f)}\ \text{Re}\int_{-\infty}^{\infty}\partial_i z_\alpha(t)\partial_j z_\alpha^*(t)dt,
\end{split}
\end{equation}
where $\alpha$ represents the TDI channel, $\partial_i z_\alpha=\partial z_\alpha/\partial \bm{\Lambda}_i$ and $\bm{\Lambda}_i$ is the $i$th parameter of GW source.
For monochromatic sources with the frequency $f$, because there is almost no frequency evolution, by using Parseval's theorem \cite{Cutler:1997ta, Vecchio:2004ec} the noise function $P_{n}^\alpha(f)$ can be taken out from the integration.
The covariance matrix $\sigma_{ij}$ between the parameters errors $\Delta\bm{\Lambda}_i=\bm{\Lambda}_i-\langle\bm{\Lambda}_i\rangle$ and $\Delta\bm{\Lambda}_j$ can be approximated by the inverse of the Fisher matrix in the large SNR limit,
\begin{equation}
\sigma_{ij}=\left\langle\Delta\bm{\Lambda}_i\Delta\bm{\Lambda}_j\right\rangle\approx (\Gamma^{-1})_{ij}.
\end{equation}
The angular uncertainty of the sky localization is defined as
\begin{equation}
\Delta \Omega_s = 2\pi\left|\sin\theta\right|
\sqrt{\sigma_{\theta\theta}\sigma_{\phi\phi}-\sigma^2_{\theta\phi}}.
\end{equation}

For a network of $n$ detectors, the SNR and FIM are defined as $\rho^2=\Sigma^{n}_{\alpha=1}\rho_{\alpha}^2$ and $\Gamma_{ij}=\Sigma_{\alpha=1}^{n}\Gamma^{\alpha}_{ij}$,
respectively.

\section{Results}\label{result}
In this section, we show the results of angular resolutions with different fiducial detectors using six TDI combinations
and discuss the impacts of different TDI combinations and detector constellations on the sky localization for monochromatic sources.

We simulate 3600 GW sources with six parameters $\bm{\Lambda} = \left(\theta,\ \phi,\ \mathcal{A},\ \iota,\ \psi,\ \phi_0\right)$, where $\theta$ and $\phi$ are chosen randomly in $\left[-\pi/2,\pi/2 \right]$ and $\left[-\pi,\pi \right]$, $\iota=1$ radian, $\psi=\phi_0=0$.
Following Ref. \cite{Zhang:2020hyx}, we fix the amplitude $\mathcal{A}$ by considering sources with the same masses and distance randomly distributed in the sky.
We take the sources to be equal
mass binary system with the total mass $(6,3\times10^2,10^4)M_\odot$ at the distance $(2.3,1.3\times10^3,10^4)$ Mpc so that the minimum SNR $\rho>10$, the corresponding GW frequencies are  $(10^{-1},10^{-2},10^{-3})$ Hz, respectively.
The benchmark frequencies $f_0=1$ mHz, $f_0=0.01$ Hz and $f_0=0.1$ Hz represent the relative low, medium and high frequency regions.

The mean and median values of the angular resolutions without TDI are shown in table \ref{tdivalue}.
In figure \ref{ratio}, we show the mean as well as the $1\sigma$ values of the ratio of the angular resolutions with TDI combinations to those
without TDI for the fiducial detectors R and C1.
The mean and the $1\sigma$ values of the ratio of the angular resolutions between two fiducial detectors are shown in figure \ref{CRC1R1}.
We also plot the ratio of the sensitivity curves between the two fiducial detectors R1 and R with and without TDI at the three benchmark frequencies in figure \ref{CRC1R1}.
The sensitivity curves of LISA, Taiji and TianQin with the Michelson $(X,Y,Z)$ combination are shown in figure \ref{sn2}.
For space-based GW detectors LISA, Taiji and TianQin and the LISA-Taiji-TianQin network with the Michelson $X$ combination, 
the mean and median values of the angular resolutions are shown in table \ref{netvalue}, the sky maps of the angular resolutions are shown in figure \ref{mapnetwork}.

\begin{table}
\centering
\begin{tabular}{ p{1.2cm}|c c c c c }
    \hline\hline
    &\multicolumn{5}{c}{The mean value of $\Delta\Omega_s$}\\
    \hline
    $f\ (\text{Hz})$  & $\text{R}$ & $\text{C}$  &$\text{R1}$  & $\text{C1}$ &$\text{C2}$\\
    \hline
    $0.001$ 
    & $1.0920\times10^{-3}$ & $2.4432\times10^{-3}$  &$5.1626\times10^{-1}$  & $1.1256$ &$1.1293$\\
    \hline
    $0.01$ 
     & $1.9150\times10^{-5}$ & $2.1554\times10^{-5}$  &$7.4818\times10^{-3}$  & $8.0221\times10^{-3}$ &$8.0474\times10^{-3}$\\
    \hline
    $0.1$  
    & $2.6030\times10^{-7}$ & $2.7312\times10^{-7}$  &$5.9204\times10^{-6}$  & $5.5359\times10^{-6}$ &$5.5201\times10^{-6}$\\
    \hline\hline
    &\multicolumn{5}{c}{The median value of $\Delta\Omega_s$}\\
    \hline
    $f\ (\text{Hz})$  & $\text{R}$ & $\text{C}$  &$\text{R1}$  & $\text{C1}$ &$\text{C2}$\\
    \hline
    $0.001$ 
    & $1.0996\times10^{-3}$ & $1.3618\times10^{-3}$  &$5.1946\times10^{-1}$  & $6.3687\times10^{-1}$ &$6.3901\times10^{-1}$\\
    \hline
    $0.01$  
    & $1.3494\times10^{-5}$ & $1.1919\times10^{-5}$  &$5.4821\times10^{-3}$  & $4.5385\times10^{-3}$ &$4.5536\times10^{-3}$\\
    \hline
    $0.1$  
    & $1.3888\times10^{-7}$ & $1.4228\times10^{-7}$  &$3.7552\times10^{-6}$  & $3.1334\times10^{-6}$ &$3.1142\times10^{-6}$\\
    \hline\hline
\end{tabular}
\caption{The mean and median values of the angular resolutions $\Delta\Omega_s$ for different fiducial detectors without TDI.}
\label{tdivalue}
\end{table}

\subsection{The effects of TDI combinations on sky localization}

\begin{figure}[htbp]
    \centering
    \includegraphics[width=0.99\columnwidth]{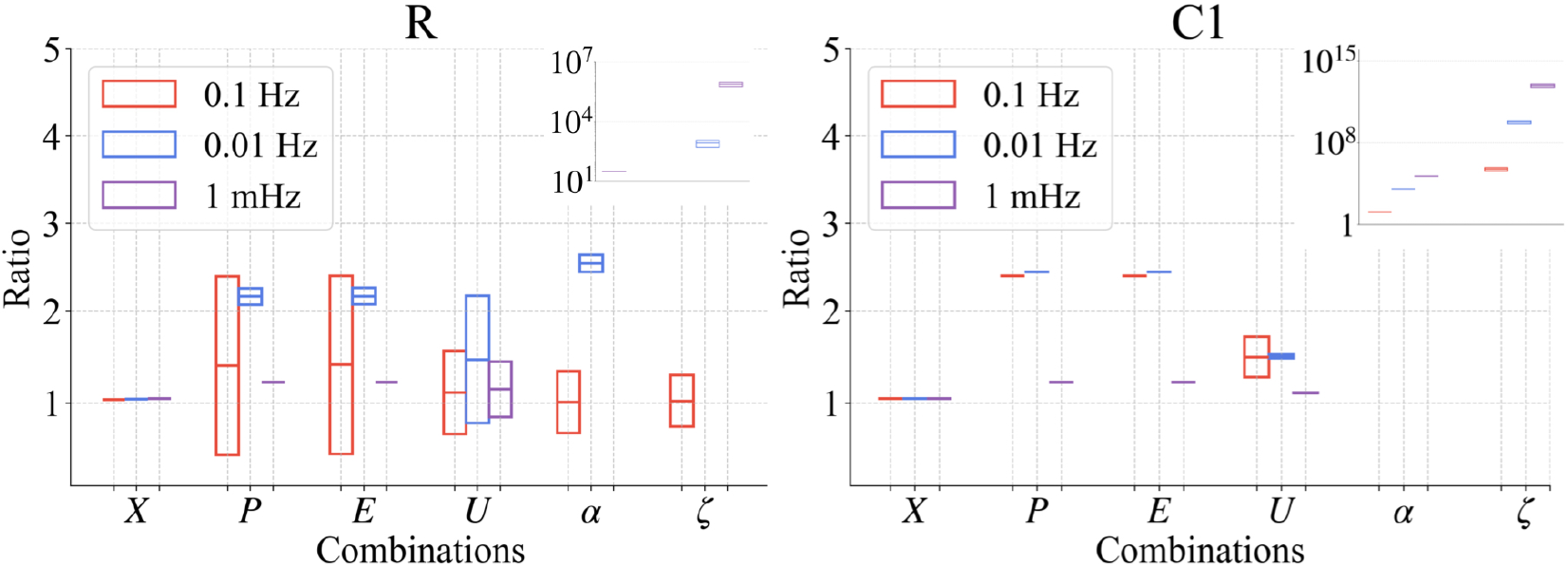}
    \caption{The mean and the $1\sigma$ values of the ratio of the angular resolutions with TDI combinations to those without TDI for the fuducial detectors R and C1.
    The colored boxes and the horizontal lines represent 1$\sigma$ and the mean values of the ratio of the angular resolutions, respectively.
    The left panel is for the fiducial detector R and the right panel is for the fiducial detector C1.
    The angular resolutions are worse if the ratio is larger than 1. 
    The inserts show the huge ratios with $\alpha$ and $\zeta$ combinations.}
    \label{ratio}
\end{figure}

To explore the effects of TDI combinations on the sky localization, 
we calculate the angular resolutions of different fiducial detectors with and without TDI.
The mean and median values of the angular resolutions without TDI are shown in table \ref{tdivalue}.
The results without TDI can be regarded as a baseline for comparison.
The mean and the $1\sigma$ values of the ratio between the angular resolutions with and without TDI for the fiducial detectors R and C1 are shown in figure \ref{ratio}. 
Since the results for the other fiducial detectors are similar, we don't show them in figure \ref{ratio}. 
As shown in figure \ref{sensitivity}, for the fiducial detectors R and C,
the $(X, Y, Z)$ combination has better sensitivities at medium and low frequencies, 
we expect better sky localization with the $(X, Y, Z)$ combination than the other TDI combinations. 
At the low frequency $f_0=1$ mHz, the sensitivity $S_n(f)$ with $\alpha$ and $\zeta$ combinations is larger by a few orders of magnitude, 
so the angular resolutions are much worse.
In fact, for the choices of parameters given above, the sources cannot be detected with $\alpha$ and $\zeta$ combinations because their SNRs are too small.
At the frequency $f_0=0.1$ Hz, the difference between the sensitivities for the six TDI combinations is less than $5\%$, 
so difference between the angular resolutions with different TDI combinations is also less than $5\%$. 
The results for the angular resolutions with the fiducial detectors R and C are consistent with these expectations.
For the fiducial detectors R1, C1 and C2,
the $(X,Y,Z)$ combination has better sensitivities at frequencies $f_0\le 0.1$ Hz, 
so the sky localization with the $(X,Y,Z)$ combination is better than the other TDI combinations as shown in figure \ref{ratio}.

For the fiducial detector R,
at $f_0=0.1$ Hz, the ratios between the mean values of the angular resolutions with and without TDI for all TDI combinations are no more than 1.39;
at $f_0=0.01$ Hz, the ratios for the $X$, $P$, $E$, $U$ and $\alpha$ combinations are less than 2.55,
but the ratio reaches $8.8 \times 10^2$ for the $\zeta$ combination;
at $f_0=1$ mHz, the ratios for the $X$, $P$, $E$, and $U$ combinations are less than 2.55,
the ratios for the $\alpha$ and $\zeta$ combinations are 32 and $7.4 \times 10^5$, respectively.
For the fiducial detector C1,
the ratios for the $X$, $P$, $E$, and $U$ combinations are all within 2.17,
but the ratios for the $\alpha$ and $\zeta$ combinations are much larger.
In fact, the ratio between the mean values of the angular resolutions with and without TDI for all TDI combinations equals to the ratio of the sensitivities $S_n^{\text{TDI}}/S_n^{\text{no-TDI}}$.
The results explain why the sky localization with the $(X,Y,Z)$ combination is better.
As shown in figure \ref{ratio}, the sky localization with the Sagnac combination $\zeta$ is much worse
because its response to GWs becomes of higher order \cite{Armstrong_1999}, 
so we don't consider the $\zeta$ combination in the discussion below.

\begin{figure}[htbp]
    \centering
    \includegraphics[width=0.99\columnwidth]{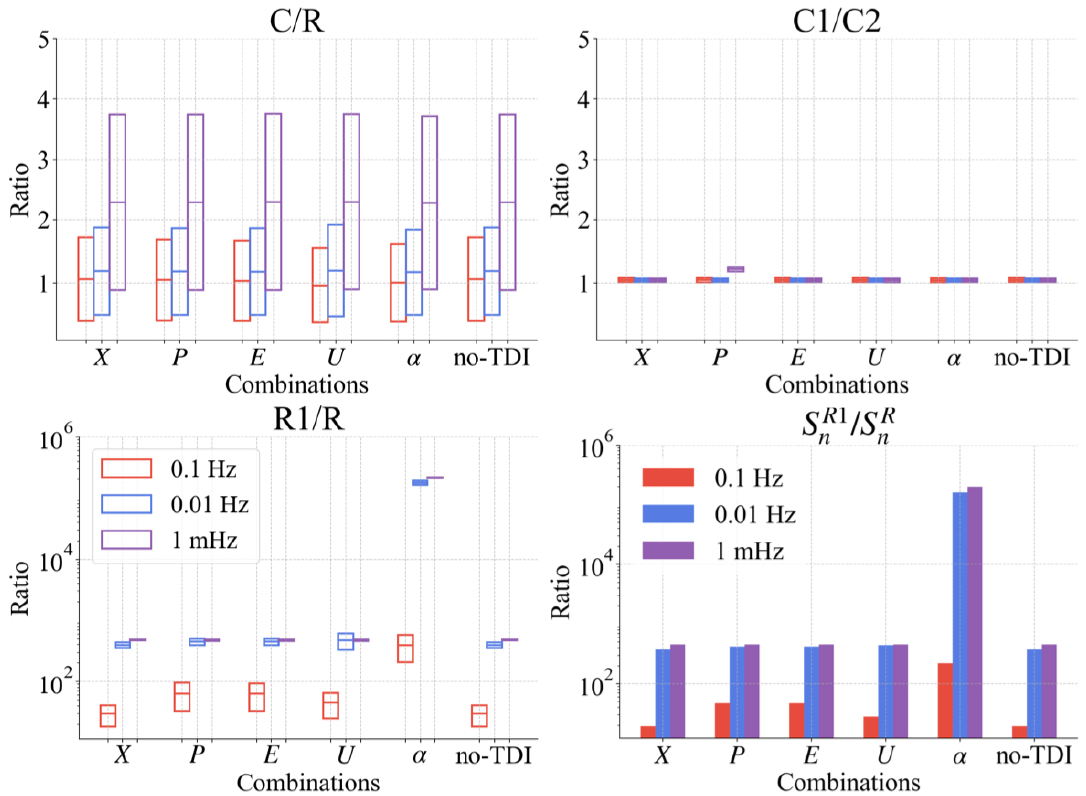}
    \caption{The mean and the $1\sigma$ values of the ratio of the angular resolutions between two fiducial detectors with different TDI combinations.
   The  colored boxes and the horizontal lines represent 1$\sigma$ and the mean values of the ratio of the angular resolutions, respectively.
    The ratio of the sensitivity curves between the fiducial detectors R and R1 with and without TDI at the three benchmark frequencies is shown in the bottom right panel.}
    \label{CRC1R1}
\end{figure}

To explore the effects of the constellations on the sky localization with different TDI combinations, 
we show the mean and $1\sigma$ values of the ratios of the angular resolutions between the fiducial detectors R and C, 
C1 and C2, R1 and R for different TDI combinations in figure \ref{CRC1R1}.
Comparing the results between R and C as shown in the upper left panel of figure \ref{CRC1R1},
we see that the rotation of the normal vector to the detector plane helps improve the mean value of the angular resolutions by a factor $\sim 2$ at $f_0=1$ mHz and the effect of the detector's changing orientation is negligible at $f_0=0.01$ Hz and $f_0=0.1$ Hz. 
The results for the effects of the rotation of the normal vector to the detector plane on the angular resolutions are the same as those found in Ref. \cite{Zhang:2020hyx}.
The effect of the detector's changing orientation on the sky localization is similar regardless of TDI combinations (the difference between different TDI combinations is less than $8\%$ at $f_0=1$ mHz, $2\%$ at $f_0=0.01$ Hz and $f_0=0.1$ Hz.), 
i.e., the ratio of the mean values of the angular resolutions between the fiducial detectors C and R is approximately 2.24 at $f_0=1$ mHz, 1.1 at $f_0=0.01$ Hz and $f_0=0.1$ Hz for all TDI combinations and no-TDI.
To see the effect of the orbital period, following Ref. \cite{Zhang:2020hyx} we compare the fiducial detectors C1 and C2
and the results  are shown in the upper right panel of figure \ref{CRC1R1}. 
As found in \cite{Zhang:2020hyx}, the influence of the rotation period of the spacecraft for the sky localization is almost negligible.
The ratio of the angular resolutions between the fiducial detectors C1 and C2 is approximately 1 at the three benchmark frequencies for all TDI combinations, 
the difference between different TDI combinations is less than $1\%$ at all three frequencies except the $P$ combination which is about $23.7\%$ at $f_0=1$ mHz.

The arm length affects both the noise and 
the transfer function of the detectors, 
so the angular resolutions for detectors with different arm lengths depend on the sensitivity curve. 
For better understanding of the result,
we also show the ratios of the sensitivity curves between the fiducial detectors R and R1 with and without TDI at the three benchmark frequencies in figure \ref{CRC1R1}.
The results in figure  \ref{CRC1R1} show that with detectors 1 and 2, the ratio between the angular resolutions $\Delta\Omega_{s,1} / \Delta\Omega_{s,2}=S_{n,1}/S_{n,2}$ for all TDI combinations.
Since the influence of different TDI combination on the sky localization  mainly comes from the sensitivity curve of the detector with the TDI combination and the $(X,Y,Z)$ combination has better sensitivity,
therefore we consider the $(X,Y,Z)$ combination only in the next section.

\subsection{The effect of network}
\begin{figure}
    \centering
    \includegraphics[width=0.95\columnwidth]{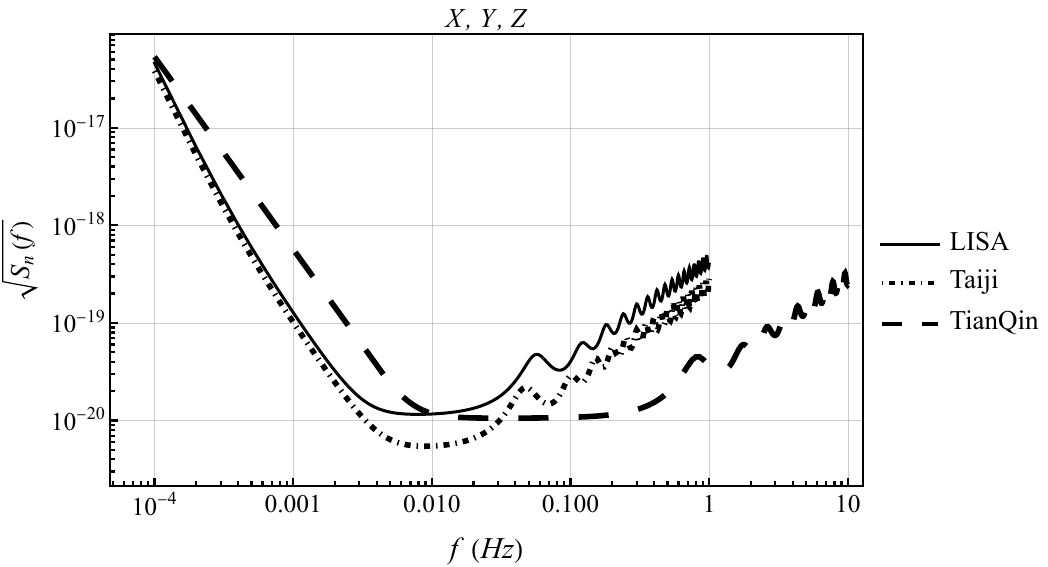}
    \caption{The sensitivity curves of LISA, Taiji and TianQin with the Michelson $(X,Y,Z)$ combination.}
    \label{sn2}
\end{figure}

\begin{figure}
    \centering
    \includegraphics[width=0.95\columnwidth]{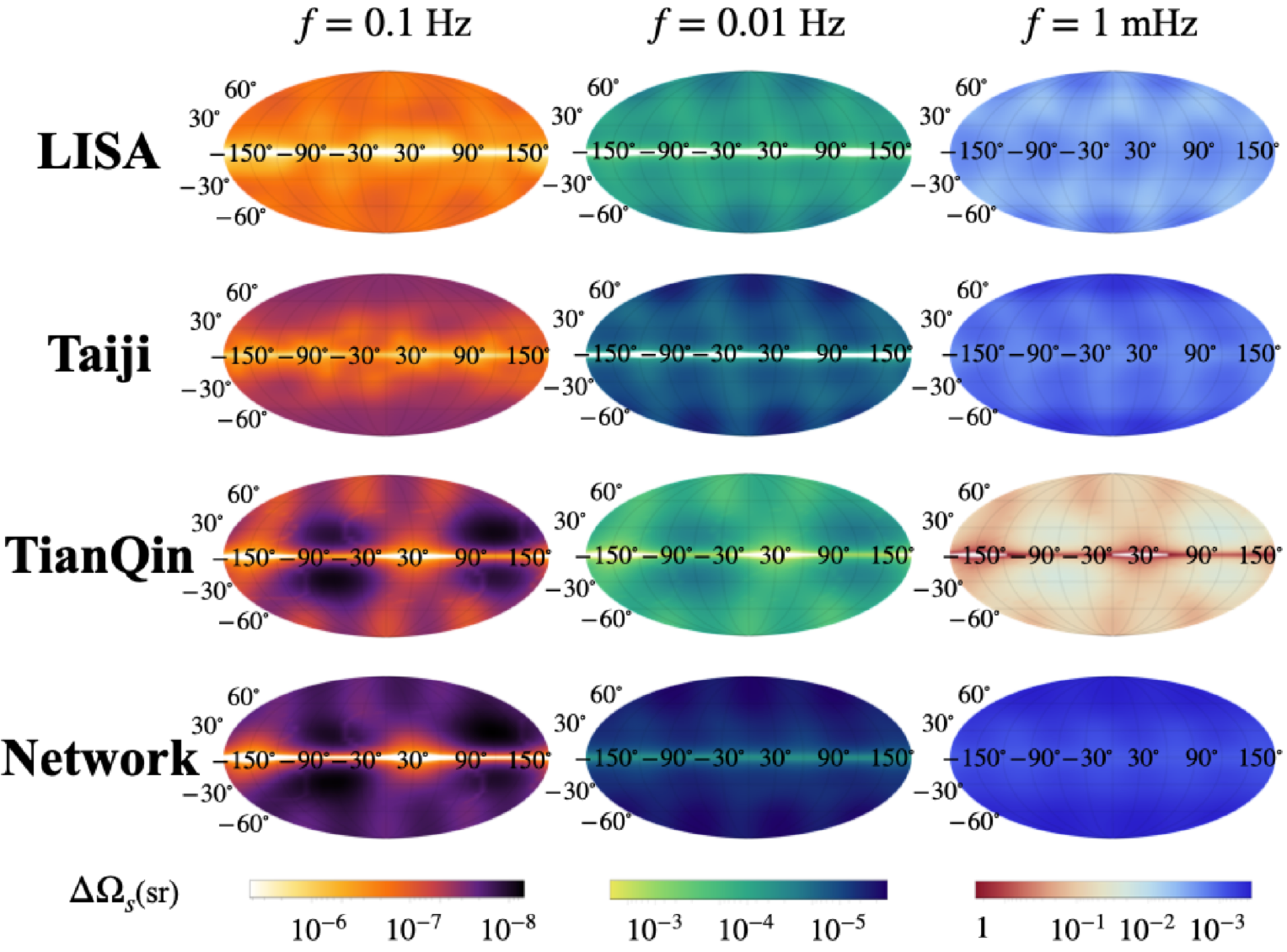}
    \caption{The sky map of angular resolutions in the unit of steradian for LISA, Taiji, TianQin, and the LISA-Taiji-TianQin network with the TDI $(X, Y, Z)$  combination. From left to right, the frequencies are $0.1$ Hz, $0.01$ Hz and $1$ mHz.}
    \label{mapnetwork}
\end{figure}

In this subsection, we apply the TDI $(X,Y,Z)$ combination to discuss the ability of sky localization for space-based GW detectors LISA, Taiji and TianQin and the LISA-Taiji-TianQin network.
The sensitivity curves of LISA, Taiji and TianQin with the Michelson $(X,Y,Z)$ combination are shown in figure \ref{sn2}.
The mean and median values of angular resolutions are shown in table \ref{netvalue}.
In table \ref{netvalue} we also show the mean and median values of angular resolutions for space-based GW detectors and their network without applying the TDI method.
The sky map of angular resolutions is shown in figure \ref{mapnetwork}.

\begin{table}
\centering
\begin{tabular}{ p{1.2cm}|c c c c}
    \hline\hline
    &\multicolumn{4}{c}{The mean value of $\Delta\Omega_s$}\\
    \hline
    $f\ (\text{Hz})$  & $\text{LISA}$ & $\text{Taiji}$  &$\text{TianQin}$  & $\text{Network}$ \\
    \hline
    $0.001$  & $2.6849\times10^{-3}$ & $1.3030\times10^{-3}$  &$9.1255\times10^{-2}$  & $ 6.1312\times10^{-4}$ \\
    & $(2.6881\times10^{-3})$ & $(1.3055\times10^{-3})$  &$(9.1256\times10^{-2})$  & $(6.1410\times10^{-4})$ \\
    \hline
    $0.01$  & $3.6733\times10^{-5}$ & $7.8631\times10^{-6}$  &$4.1719\times10^{-5}$  & $4.3272\times10^{-6}$ \\
    & $(3.6890\times10^{-5})$ & $(8.0162\times10^{-6})$  &$(4.1725\times10^{-5})$  & $(4.3838\times10^{-6})$ \\
    \hline
    $0.1$ & $3.0966\times10^{-7}$ & $1.5947\times10^{-7}$  &$2.3068\times10^{-8}$  & $1.6795\times10^{-8}$ \\
    & $(3.1250\times10^{-7})$ & $(1.5947\times10^{-7})$  &($2.3068\times10^{-8})$  & $(1.6808\times10^{-8})$ \\
    \hline\hline
    &\multicolumn{4}{c}{The median value of $\Delta\Omega_s$}\\ 
    \hline
    $f\ (\text{Hz})$  & $\text{LISA}$ & $\text{Taiji}$  &$\text{TianQin}$  & $\text{Network}$ \\
    \hline
    $0.001$  & $2.5871\times10^{-3}$ & $1.3051\times10^{-3}$  &$5.1638\times10^{-2}$  & $6.0841\times10^{-4}$ \\
    & $(2.59021\times10^{-3})$ & $(1.3076\times10^{-3})$  &$(5.1638\times10^{-2})$  & $(6.0939\times10^{-4})$ \\
    \hline
    $0.01$  & $2.7778\times10^{-5}$ & $5.7658\times10^{-6}$  &$2.3607\times10^{-5}$  & $2.8500\times10^{-6}$ \\ 
    & $(2.79001\times10^{-5})$ & $(5.8981\times10^{-6})$  &$(2.3610\times10^{-5})$  & $(2.8856\times10^{-6})$ \\
    \hline
    $0.1$ & $1.6370\times10^{-7}$ & $7.1032\times10^{-8}$  &$1.3014\times10^{-8}$  & $8.8234\times10^{-9}$ \\
    & $(1.6521\times10^{-7})$ & $(7.1032\times10^{-8})$  &$(1.3014\times10^{-8})$  & $(8.8313\times10^{-9})$ \\
    \hline\hline
\end{tabular}
\caption{The mean and median values of angular resolutions $\Delta\Omega_s$ for space-based GW detectors and their network with Michelson $(X, Y, Z)$ combination and without TDI, the results in brackets are for those without TDI.}
\label{netvalue}
\end{table}

At the frequency $f_0=1$ mHz, the amplitude modulation helps LISA and Taiji on improving the sky localization and there is no equatorial pattern in the sky map.
For the sky localization, LISA is better than TianQin by about 34 times,
Taiji is two times better than LISA,
and the LISA-Taiji-TianQin network is $2.5$ times better than Taiji.
At the relative medium and high frequencies $f_0=0.01$ Hz and $f_0=0.1$ Hz,
the amplitude modulation is negligible and the Doppler modulation dominates,
the equatorial pattern exists in the sky map as shown in figure \ref{mapnetwork}.

At the medium frequency $f_0=0.01$ Hz,
the GW wavelength is still bigger than the arm length of LISA and Taiji;
for the sky localization, LISA is less than two times better than TianQin,
Taiji is almost five times better than LISA,
and the LISA-Taiji-TianQin network is less than two times better than Taiji.

As the frequency of GWs increases, the wavelength of GWs becomes shorter. 
When the wavelength of GWs is shorter than detector's arm length, 
the frequency-dependent transfer function can deteriorate the SNR registered in the detector.
At the relatively high frequency $f_0=0.1$ Hz,
the GW wavelength is smaller than the arm length of LISA and Taiji,
but bigger than the arm length of TianQin,
so TianQin has a better ability of sky localization than LISA and Taiji. 
For the sky localization, 
Taiji is two times better than LISA,
TianQin is seven times better than Taiji,
and the LISA-Taiji-TianQin network is around ten times better than Taiji.

For monochromatic sources, the detector in different orbital positions can be thought as an independent detector, 
so the improvement on the angular resolutions by the network of combined detectors is small.

\section{Conclusion}
\label{discussion}

For spaced-based GW detectors, the technique of TDI is needed to cancel out laser frequency noises due to the difficulty of maintaining the exact equality of the arm length in space.
We extend the previous work on the analysis of the main factors that influence the accuracy of source localization by considering six particular TDI combinations and discussing their effects on the sky localization. 

With the same detector, the ratios between the mean values of the angular resolutions with and without TDI for different TDI combinations equal to $S_n^\text{TDI}/S_n^\text{no-TDI}$.
For the effect of the detector's changing orientation on the sky localization with different TDI combinations, 
we find that the ratio of the mean values of the angular resolutions between the fiducial detectors C and R is approximately 2.24 at $f_0=1$ mHz, 1.1 at $f_0=0.01$ Hz and $f_0=0.1$ Hz for all TDI combinations including no-TDI.
For the effect of the orbital period on the sky localization with different TDI combinations, we find that the ratio of the mean values of the angular resolutions between the fiducial detectors C1 and C2  is approximately 1 at the three frequencies for all TDI combinations. 
The effect of the arm length is reflected in the sensitivity curve through the noise and the transfer function,
we find that the ratio between the mean value of the angular resolutions with detectors 1 and 2 is
$\Delta\Omega_{s,1} /\Delta\Omega_{s,2}=S_{n,1}/S_{n,2}$ for all TDI combinations.
Therefore, we conclude that the influence of different TDI combination on the sky localization mainly comes from the sensitivity curve of the detector with the TDI combination.

Since the sensitivity with the $(X,Y,Z)$ combination is better,
we apply the $(X, Y, Z)$ combination to analyze the uncertainty of sky localization for LISA, Taiji, TianQin and the LISA-Taiji-TianQin network.
At the low frequency $f_0=1$ mHz, 
the mean value of the angular resolutions with LISA is better than that with TianQin by about 34 times of magnitude with the help of the noise curve,
Taiji is two times better than LISA,
and the LISA-Taiji-TianQin network is $2.5$ times better than Taiji.
At the medium frequency $f_0=0.01$ Hz,
the mean value of the angular resolutions with LISA is less than two times better than that with TianQin,
Taiji is almost five times better than LISA,
and the LISA-Taiji-TianQin network is less than two times better than Taiji.
At the relatively high frequency $f_0=0.1$ Hz,
the mean value of the angular resolutions with Taiji is two times better than that with LISA,
TianQin is seven times better than Taiji,
and the LISA-Taiji-TianQin network is around ten times better than Taiji.

For monochromatic sources, the detector in different orbital positions can be thought as an independent detector,
so the improvement on the angular resolutions by the network of combined detectors is small.
The mean values of angular resolutions with the LISA-Taiji-TianQin network are about $6.1\times10^{-4}$ steradian at 1 mHz, $4.3\times10^{-6}$ steradian at 10 mHz and $1.7\times10^{-8}$ steradian at 100 mHz.
The difference between the angular resolutions with the $(X,Y,Z)$ combination and without TDI is less than $14\%$.

This analysis is based on the first-generation TDI with constant arm length,
the effect of time-changing arm length can be dealt with the second-generation TDI combinations
and this will be our future study.
Furthermore, we use the uncorrelated noise model in this paper to simplify the calculation, the realistic noise model for space-based GW detectors is complicated, and it should be carefully studied in the future.

\begin{acknowledgments}
This research is supported in part by the National Key Research and Development Program of China under Grant No. 2020YFC2201504 and the National Natural Science Foundation of China (NSFC) under Grant NO.12465013.
\end{acknowledgments}


\begin{thebibliography}{10}

\bibitem{LIGOScientific:2016aoc}
{\scshape LIGO Scientific, Virgo} collaboration, \emph{{Observation of
  Gravitational Waves from a Binary Black Hole Merger}},
  \href{https://doi.org/10.1103/PhysRevLett.116.061102}{\emph{Phys. Rev. Lett.}
  {\bfseries 116} (2016) 061102}
  [\href{https://arxiv.org/abs/1602.03837}{{\ttfamily 1602.03837}}].

\bibitem{Harry:2010zz}
{\scshape LIGO Scientific} collaboration, \emph{{Advanced LIGO: The next
  generation of gravitational wave detectors}},
  \href{https://doi.org/10.1088/0264-9381/27/8/084006}{\emph{Class. Quant.
  Grav.} {\bfseries 27} (2010) 084006}.

\bibitem{LIGOScientific:2014pky}
{\scshape LIGO Scientific} collaboration, \emph{{Advanced LIGO}},
  \href{https://doi.org/10.1088/0264-9381/32/7/074001}{\emph{Class. Quant.
  Grav.} {\bfseries 32} (2015) 074001}
  [\href{https://arxiv.org/abs/1411.4547}{{\ttfamily 1411.4547}}].

\bibitem{VIRGO:2014yos}
{\scshape VIRGO} collaboration, \emph{{Advanced Virgo: a second-generation
  interferometric gravitational wave detector}},
  \href{https://doi.org/10.1088/0264-9381/32/2/024001}{\emph{Class. Quant.
  Grav.} {\bfseries 32} (2015) 024001}
  [\href{https://arxiv.org/abs/1408.3978}{{\ttfamily 1408.3978}}].

\bibitem{LIGOScientific:2016lio}
{\scshape LIGO Scientific, Virgo} collaboration, \emph{{Tests of general
  relativity with GW150914}},
  \href{https://doi.org/10.1103/PhysRevLett.116.221101}{\emph{Phys. Rev. Lett.}
  {\bfseries 116} (2016) 221101}
  [\href{https://arxiv.org/abs/1602.03841}{{\ttfamily 1602.03841}}].

\bibitem{LIGOScientific:2018dkp}
{\scshape LIGO Scientific, Virgo} collaboration, \emph{{Tests of General
  Relativity with GW170817}},
  \href{https://doi.org/10.1103/PhysRevLett.123.011102}{\emph{Phys. Rev. Lett.}
  {\bfseries 123} (2019) 011102}
  [\href{https://arxiv.org/abs/1811.00364}{{\ttfamily 1811.00364}}].

\bibitem{LIGOScientific:2019fpa}
{\scshape LIGO Scientific, Virgo} collaboration, \emph{{Tests of General
  Relativity with the Binary Black Hole Signals from the LIGO-Virgo Catalog
  GWTC-1}}, \href{https://doi.org/10.1103/PhysRevD.100.104036}{\emph{Phys. Rev.
  D} {\bfseries 100} (2019) 104036}
  [\href{https://arxiv.org/abs/1903.04467}{{\ttfamily 1903.04467}}].

\bibitem{LIGOScientific:2020tif}
{\scshape LIGO Scientific, Virgo} collaboration, \emph{{Tests of general
  relativity with binary black holes from the second LIGO-Virgo
  gravitational-wave transient catalog}},
  \href{https://doi.org/10.1103/PhysRevD.103.122002}{\emph{Phys. Rev. D}
  {\bfseries 103} (2021) 122002}
  [\href{https://arxiv.org/abs/2010.14529}{{\ttfamily 2010.14529}}].

\bibitem{LIGOScientific:2016dsl}
{\scshape LIGO Scientific, Virgo} collaboration, \emph{{Binary Black Hole
  Mergers in the first Advanced LIGO Observing Run}},
  \href{https://doi.org/10.1103/PhysRevX.6.041015}{\emph{Phys. Rev. X}
  {\bfseries 6} (2016) 041015}
  [\href{https://arxiv.org/abs/1606.04856}{{\ttfamily 1606.04856}}].

\bibitem{LIGOScientific:2016kwr}
{\scshape LIGO Scientific, Virgo} collaboration, \emph{{The Rate of Binary
  Black Hole Mergers Inferred from Advanced LIGO Observations Surrounding
  GW150914}},
  \href{https://doi.org/10.3847/2041-8205/833/1/L1}{\emph{Astrophys. J. Lett.}
  {\bfseries 833} (2016) L1}
  [\href{https://arxiv.org/abs/1602.03842}{{\ttfamily 1602.03842}}].

\bibitem{LIGOScientific:2016hpm}
{\scshape LIGO Scientific, Virgo} collaboration, \emph{{Upper Limits on the
  Rates of Binary Neutron Star and Neutron Star\textendash{}black Hole Mergers
  From Advanced Ligo\textquoteright{}s First Observing run}},
  \href{https://doi.org/10.3847/2041-8205/832/2/L21}{\emph{Astrophys. J. Lett.}
  {\bfseries 832} (2016) L21}
  [\href{https://arxiv.org/abs/1607.07456}{{\ttfamily 1607.07456}}].

\bibitem{LIGOScientific:2021aug}
{\scshape LIGO Scientific, Virgo, KAGRA} collaboration, \emph{{Constraints on
  the Cosmic Expansion History from GWTC\textendash{}3}},
  \href{https://doi.org/10.3847/1538-4357/ac74bb}{\emph{Astrophys. J.}
  {\bfseries 949} (2023) 76}
  [\href{https://arxiv.org/abs/2111.03604}{{\ttfamily 2111.03604}}].

\bibitem{Seto:2001qf}
N.~Seto, S.~Kawamura and T.~Nakamura, \emph{{Possibility of direct measurement
  of the acceleration of the universe using 0.1-Hz band laser interferometer
  gravitational wave antenna in space}},
  \href{https://doi.org/10.1103/PhysRevLett.87.221103}{\emph{Phys. Rev. Lett.}
  {\bfseries 87} (2001) 221103}
  [\href{https://arxiv.org/abs/astro-ph/0108011}{{\ttfamily
  astro-ph/0108011}}].

\bibitem{Kyutoku:2016zxn}
K.~Kyutoku and N.~Seto, \emph{{Gravitational-wave cosmography with LISA and the
  Hubble tension}},
  \href{https://doi.org/10.1103/PhysRevD.95.083525}{\emph{Phys. Rev. D}
  {\bfseries 95} (2017) 083525}
  [\href{https://arxiv.org/abs/1609.07142}{{\ttfamily 1609.07142}}].

\bibitem{eLISA:2013xep}
{\scshape eLISA} collaboration, \emph{{The Gravitational Universe}},
  \href{https://arxiv.org/abs/1305.5720}{{\ttfamily 1305.5720}}.

\bibitem{Klein:2015hvg}
A.~Klein et~al., \emph{{Science with the space-based interferometer eLISA:
  Supermassive black hole binaries}},
  \href{https://doi.org/10.1103/PhysRevD.93.024003}{\emph{Phys. Rev. D}
  {\bfseries 93} (2016) 024003}
  [\href{https://arxiv.org/abs/1511.05581}{{\ttfamily 1511.05581}}].

\bibitem{Sesana:2008ur}
A.~Sesana, M.~Volonteri and F.~Haardt, \emph{{LISA detection of massive black
  hole binaries: imprint of seed populations and of exterme recoils}},
  \href{https://doi.org/10.1088/0264-9381/26/9/094033}{\emph{Class. Quant.
  Grav.} {\bfseries 26} (2009) 094033}
  [\href{https://arxiv.org/abs/0810.5554}{{\ttfamily 0810.5554}}].

\bibitem{Ricarte:2018mzn}
A.~Ricarte and P.~Natarajan, \emph{{The Observational Signatures of
  Supermassive Black Hole Seeds}},
  \href{https://doi.org/10.1093/mnras/sty2448}{\emph{Mon. Not. Roy. Astron.
  Soc.} {\bfseries 481} (2018) 3278}
  [\href{https://arxiv.org/abs/1809.01177}{{\ttfamily 1809.01177}}].

\bibitem{Li:2022fno}
K.~Li, T.~Bogdanovi\'c, D.~R. Ballantyne and M.~Bonetti, \emph{{Massive Black
  Hole Binaries from the TNG50-3 Simulation. I. Coalescence and LISA Detection
  Rates}}, \href{https://doi.org/10.3847/1538-4357/ac74b5}{\emph{Astrophys. J.}
  {\bfseries 933} (2022) 104}
  [\href{https://arxiv.org/abs/2201.11088}{{\ttfamily 2201.11088}}].

\bibitem{Mangiagli:2022niy}
A.~Mangiagli, C.~Caprini, M.~Volonteri, S.~Marsat, S.~Vergani, N.~Tamanini
  et~al., \emph{{Massive black hole binaries in LISA: Multimessenger prospects
  and electromagnetic counterparts}},
  \href{https://doi.org/10.1103/PhysRevD.106.103017}{\emph{Phys. Rev. D}
  {\bfseries 106} (2022) 103017}
  [\href{https://arxiv.org/abs/2207.10678}{{\ttfamily 2207.10678}}].

\bibitem{Danzmann:1997hm}
K.~Danzmann, \emph{{LISA: An ESA cornerstone mission for a gravitational wave
  observatory}}, \href{https://doi.org/10.1088/0264-9381/14/6/002}{\emph{Class.
  Quant. Grav.} {\bfseries 14} (1997) 1399}.

\bibitem{LISA:2017pwj}
{\scshape LISA} collaboration, \emph{{Laser Interferometer Space Antenna}},
  \href{https://arxiv.org/abs/1702.00786}{{\ttfamily 1702.00786}}.

\bibitem{Hu:2017mde}
W.-R. Hu and Y.-L. Wu, \emph{{The Taiji Program in Space for gravitational wave
  physics and the nature of gravity}},
  \href{https://doi.org/10.1093/nsr/nwx116}{\emph{Natl. Sci. Rev.} {\bfseries
  4} (2017) 685}.

\bibitem{TianQin:2015yph}
{\scshape TianQin} collaboration, \emph{{TianQin: a space-borne gravitational
  wave detector}},
  \href{https://doi.org/10.1088/0264-9381/33/3/035010}{\emph{Class. Quant.
  Grav.} {\bfseries 33} (2016) 035010}
  [\href{https://arxiv.org/abs/1512.02076}{{\ttfamily 1512.02076}}].

\bibitem{Grover:2013sha}
K.~Grover, S.~Fairhurst, B.~F. Farr, I.~Mandel, C.~Rodriguez, T.~Sidery et~al.,
  \emph{{Comparison of Gravitational Wave Detector Network Sky Localization
  Approximations}},
  \href{https://doi.org/10.1103/PhysRevD.89.042004}{\emph{Phys. Rev. D}
  {\bfseries 89} (2014) 042004}
  [\href{https://arxiv.org/abs/1310.7454}{{\ttfamily 1310.7454}}].

\bibitem{Cutler:1997ta}
C.~Cutler, \emph{{Angular resolution of the LISA gravitational wave detector}},
  \href{https://doi.org/10.1103/PhysRevD.57.7089}{\emph{Phys. Rev. D}
  {\bfseries 57} (1998) 7089}
  [\href{https://arxiv.org/abs/gr-qc/9703068}{{\ttfamily gr-qc/9703068}}].

\bibitem{Berry:2014jja}
C.~P.~L. Berry et~al., \emph{{Parameter estimation for binary neutron-star
  coalescences with realistic noise during the Advanced LIGO era}},
  \href{https://doi.org/10.1088/0004-637X/804/2/114}{\emph{Astrophys. J.}
  {\bfseries 804} (2015) 114}
  [\href{https://arxiv.org/abs/1411.6934}{{\ttfamily 1411.6934}}].

\bibitem{Schutz:1986gp}
B.~F. Schutz, \emph{{Determining the Hubble Constant from Gravitational Wave
  Observations}}, \href{https://doi.org/10.1038/323310a0}{\emph{Nature}
  {\bfseries 323} (1986) 310}.

\bibitem{Holz:2005df}
D.~E. Holz and S.~A. Hughes, \emph{{Using gravitational-wave standard sirens}},
  \href{https://doi.org/10.1086/431341}{\emph{Astrophys. J.} {\bfseries 629}
  (2005) 15} [\href{https://arxiv.org/abs/astro-ph/0504616}{{\ttfamily
  astro-ph/0504616}}].

\bibitem{Chen:2017rfc}
H.-Y. Chen, M.~Fishbach and D.~E. Holz, \emph{{A two per cent Hubble constant
  measurement from standard sirens within five years}},
  \href{https://doi.org/10.1038/s41586-018-0606-0}{\emph{Nature} {\bfseries
  562} (2018) 545} [\href{https://arxiv.org/abs/1712.06531}{{\ttfamily
  1712.06531}}].

\bibitem{Hotokezaka:2018dfi}
K.~Hotokezaka, E.~Nakar, O.~Gottlieb, S.~Nissanke, K.~Masuda, G.~Hallinan
  et~al., \emph{{A Hubble constant measurement from superluminal motion of the
  jet in GW170817}},
  \href{https://doi.org/10.1038/s41550-019-0820-1}{\emph{Nature Astron.}
  {\bfseries 3} (2019) 940} [\href{https://arxiv.org/abs/1806.10596}{{\ttfamily
  1806.10596}}].

\bibitem{Vitale:2018wlg}
S.~Vitale and H.-Y. Chen, \emph{{Measuring the Hubble constant with neutron
  star black hole mergers}},
  \href{https://doi.org/10.1103/PhysRevLett.121.021303}{\emph{Phys. Rev. Lett.}
  {\bfseries 121} (2018) 021303}
  [\href{https://arxiv.org/abs/1804.07337}{{\ttfamily 1804.07337}}].

\bibitem{LIGOScientific:2019zcs}
{\scshape LIGO Scientific, Virgo, VIRGO} collaboration, \emph{{A
  Gravitational-wave Measurement of the Hubble Constant Following the Second
  Observing Run of Advanced LIGO and Virgo}},
  \href{https://doi.org/10.3847/1538-4357/abdcb7}{\emph{Astrophys. J.}
  {\bfseries 909} (2021) 218}
  [\href{https://arxiv.org/abs/1908.06060}{{\ttfamily 1908.06060}}].

\bibitem{Zhu:2021bpp}
L.-G. Zhu, L.-H. Xie, Y.-M. Hu, S.~Liu, E.-K. Li, N.~R. Napolitano et~al.,
  \emph{{Constraining the Hubble constant to a precision of about 1\% using
  multi-band dark standard siren detections}},
  \href{https://doi.org/10.1007/s11433-021-1859-9}{\emph{Sci. China Phys. Mech.
  Astron.} {\bfseries 65} (2022) 259811}
  [\href{https://arxiv.org/abs/2110.05224}{{\ttfamily 2110.05224}}].

\bibitem{Riess:2019cxk}
A.~G. Riess, S.~Casertano, W.~Yuan, L.~M. Macri and D.~Scolnic, \emph{{Large
  Magellanic Cloud Cepheid Standards Provide a 1\% Foundation for the
  Determination of the Hubble Constant and Stronger Evidence for Physics beyond
  $\Lambda$CDM}},
  \href{https://doi.org/10.3847/1538-4357/ab1422}{\emph{Astrophys. J.}
  {\bfseries 876} (2019) 85}
  [\href{https://arxiv.org/abs/1903.07603}{{\ttfamily 1903.07603}}].

\bibitem{Blaut:2011zz}
A.~Blaut, \emph{{Accuracy of estimation of parameters with LISA}},
  \href{https://doi.org/10.1103/PhysRevD.83.083006}{\emph{Phys. Rev. D}
  {\bfseries 83} (2011) 083006}.

\bibitem{Zhang:2023ceh}
C.~Zhang, N.~Dai and D.~Liang, \emph{{Importance of including higher signal
  harmonics in the modeling of extreme mass-ratio inspirals}},
  \href{https://doi.org/10.1103/PhysRevD.108.044076}{\emph{Phys. Rev. D}
  {\bfseries 108} (2023) 044076}
  [\href{https://arxiv.org/abs/2306.13871}{{\ttfamily 2306.13871}}].

\bibitem{Zhang:2020hyx}
C.~Zhang, Y.~Gong, H.~Liu, B.~Wang and C.~Zhang, \emph{{Sky localization of
  space-based gravitational wave detectors}},
  \href{https://doi.org/10.1103/PhysRevD.103.103013}{\emph{Phys. Rev. D}
  {\bfseries 103} (2021) 103013}
  [\href{https://arxiv.org/abs/2009.03476}{{\ttfamily 2009.03476}}].

\bibitem{Zhang:2020drf}
C.~Zhang, Y.~Gong, B.~Wang and C.~Zhang, \emph{{Accuracy of parameter
  estimations with a spaceborne gravitational wave observatory}},
  \href{https://doi.org/10.1103/PhysRevD.103.104066}{\emph{Phys. Rev. D}
  {\bfseries 103} (2021) 104066}
  [\href{https://arxiv.org/abs/2012.01043}{{\ttfamily 2012.01043}}].

\bibitem{Gong:2021gvw}
Y.~Gong, J.~Luo and B.~Wang, \emph{{Concepts and status of Chinese space
  gravitational wave detection projects}},
  \href{https://doi.org/10.1038/s41550-021-01480-3}{\emph{Nature Astron.}
  {\bfseries 5} (2021) 881} [\href{https://arxiv.org/abs/2109.07442}{{\ttfamily
  2109.07442}}].

\bibitem{Peterseim:1996cw}
M.~Peterseim, O.~Jennrich and K.~Danzmann, \emph{{Accuracy of parameter
  estimation of gravitational waves with LISA}},
  \href{https://doi.org/10.1088/0264-9381/13/11A/037}{\emph{Class. Quant.
  Grav.} {\bfseries 13} (1996) A279}.

\bibitem{Tinto:1994kg}
M.~Tinto, G.~Giampieri, R.~W. Hellings, P.~L. Bender and J.~E. Faller,
  \emph{{Algorithms for unequal-arm Michelson interferometers}},  in \emph{{7th
  Marcel Grossmann Meeting on General Relativity (MG 7)}}, pp.~1668--1670, 7,
  1994.

\bibitem{Tinto:1999yr}
M.~Tinto and J.~W. Armstrong, \emph{{Cancellation of laser noise in an
  unequal-arm interferometer detector of gravitational radiation}},
  \href{https://doi.org/10.1103/PhysRevD.59.102003}{\emph{Phys. Rev. D}
  {\bfseries 59} (1999) 102003}.

\bibitem{Armstrong_1999}
J.~W. Armstrong, F.~B. Estabrook and M.~Tinto, \emph{Time-delay interferometry
  for space-based gravitational wave searches},
  \href{https://doi.org/10.1086/308110}{\emph{Astrophys. J.} {\bfseries 527}
  (1999) 814}.

\bibitem{Estabrook:2000ef}
F.~B. Estabrook, M.~Tinto and J.~W. Armstrong, \emph{{Time delay analysis of
  LISA gravitational wave data: Elimination of spacecraft motion effects}},
  \href{https://doi.org/10.1103/PhysRevD.62.042002}{\emph{Phys. Rev. D}
  {\bfseries 62} (2000) 042002}.

\bibitem{Tinto:2003vj}
M.~Tinto, F.~B. Estabrook and J.~W. Armstrong, \emph{{Time delay interferometry
  with moving spacecraft arrays}},
  \href{https://doi.org/10.1103/PhysRevD.69.082001}{\emph{Phys. Rev. D}
  {\bfseries 69} (2004) 082001}
  [\href{https://arxiv.org/abs/gr-qc/0310017}{{\ttfamily gr-qc/0310017}}].

\bibitem{Vallisneri:2004bn}
M.~Vallisneri, \emph{{Synthetic LISA: Simulating time delay interferometry in a
  model LISA}}, \href{https://doi.org/10.1103/PhysRevD.71.022001}{\emph{Phys.
  Rev. D} {\bfseries 71} (2005) 022001}
  [\href{https://arxiv.org/abs/gr-qc/0407102}{{\ttfamily gr-qc/0407102}}].

\bibitem{Shaddock:2003dj}
D.~A. Shaddock, M.~Tinto, F.~B. Estabrook and J.~W. Armstrong, \emph{{Data
  combinations accounting for LISA spacecraft motion}},
  \href{https://doi.org/10.1103/PhysRevD.68.061303}{\emph{Phys. Rev. D}
  {\bfseries 68} (2003) 061303}
  [\href{https://arxiv.org/abs/gr-qc/0307080}{{\ttfamily gr-qc/0307080}}].

\bibitem{Cornish:2003tz}
N.~J. Cornish and R.~W. Hellings, \emph{{The Effects of orbital motion on LISA
  time delay interferometry}},
  \href{https://doi.org/10.1088/0264-9381/20/22/009}{\emph{Class. Quant. Grav.}
  {\bfseries 20} (2003) 4851}
  [\href{https://arxiv.org/abs/gr-qc/0306096}{{\ttfamily gr-qc/0306096}}].

\bibitem{Tinto:2001ii}
M.~Tinto, J.~W. Armstrong and F.~B. Estabrook, \emph{{Discriminating a
  gravitational wave background from instrumental noise in the LISA detector}},
  \href{https://doi.org/10.1103/PhysRevD.63.021101}{\emph{Phys. Rev. D}
  {\bfseries 63} (2001) 021101}.

\bibitem{Tinto:2001ui}
M.~Tinto, J.~W. Armstrong and F.~B. Estabrook, \emph{{Discriminating a
  gravitational-wave background from instrumental noise using time-delay
  interferometry}},
  \href{https://doi.org/10.1088/0264-9381/18/19/316}{\emph{Class. Quant. Grav.}
  {\bfseries 18} (2001) 4081}.

\bibitem{Hogan:2001jn}
C.~J. Hogan and P.~L. Bender, \emph{{Estimating stochastic gravitational wave
  backgrounds with Sagnac calibration}},
  \href{https://doi.org/10.1103/PhysRevD.64.062002}{\emph{Phys. Rev. D}
  {\bfseries 64} (2001) 062002}
  [\href{https://arxiv.org/abs/astro-ph/0104266}{{\ttfamily
  astro-ph/0104266}}].

\bibitem{Armstrong:2001uh}
J.~W. Armstrong, F.~B. Estabrook and M.~Tinto, \emph{{Sensitivities of
  alternate LISA configurations}},
  \href{https://doi.org/10.1088/0264-9381/18/19/313}{\emph{Class. Quant. Grav.}
  {\bfseries 18} (2001) 4059}.

\bibitem{Prince:2002hp}
T.~A. Prince, M.~Tinto, S.~L. Larson and J.~W. Armstrong, \emph{{The LISA
  optimal sensitivity}},
  \href{https://doi.org/10.1103/PhysRevD.66.122002}{\emph{Phys. Rev. D}
  {\bfseries 66} (2002) 122002}
  [\href{https://arxiv.org/abs/gr-qc/0209039}{{\ttfamily gr-qc/0209039}}].

\bibitem{Tinto:2002de}
M.~Tinto, F.~B. Estabrook and J.~W. Armstrong, \emph{{Time delay interferometry
  for LISA}}, \href{https://doi.org/10.1103/PhysRevD.65.082003}{\emph{Phys.
  Rev. D} {\bfseries 65} (2002) 082003}.

\bibitem{Tinto:2003uk}
M.~Tinto, D.~A. Shaddock, J.~Sylvestre and J.~W. Armstrong,
  \emph{{Implementation of time delay interferometry for LISA}},
  \href{https://doi.org/10.1103/PhysRevD.67.122003}{\emph{Phys. Rev. D}
  {\bfseries 67} (2003) 122003}
  [\href{https://arxiv.org/abs/gr-qc/0303013}{{\ttfamily gr-qc/0303013}}].

\bibitem{Nayak:2003na}
K.~R. Nayak, S.~V. Dhurandhar, A.~Pai and J.~Y. Vinet, \emph{{Optimizing the
  directional sensitivity of LISA}},
  \href{https://doi.org/10.1103/PhysRevD.70.049901}{\emph{Phys. Rev. D}
  {\bfseries 68} (2003) 122001}
  [\href{https://arxiv.org/abs/gr-qc/0306050}{{\ttfamily gr-qc/0306050}}].

\bibitem{Tinto:2004nz}
M.~Tinto and S.~L. Larson, \emph{{The LISA time-delay interferometry
  zero-signal solution. I. Geometrical properties}},
  \href{https://doi.org/10.1103/PhysRevD.70.062002}{\emph{Phys. Rev. D}
  {\bfseries 70} (2004) 062002}
  [\href{https://arxiv.org/abs/gr-qc/0405147}{{\ttfamily gr-qc/0405147}}].

\bibitem{Romano:2006rj}
J.~D. Romano and G.~Woan, \emph{{A Principal component analysis for LISA: The
  TDI connection}},
  \href{https://doi.org/10.1103/PhysRevD.73.102001}{\emph{Phys. Rev. D}
  {\bfseries 73} (2006) 102001}
  [\href{https://arxiv.org/abs/gr-qc/0602033}{{\ttfamily gr-qc/0602033}}].

\bibitem{Zhang:2020khm}
C.~Zhang, Q.~Gao, Y.~Gong, B.~Wang, A.~J. Weinstein and C.~Zhang, \emph{{Full
  analytical formulas for frequency response of space-based gravitational wave
  detectors}}, \href{https://doi.org/10.1103/PhysRevD.101.124027}{\emph{Phys.
  Rev. D} {\bfseries 101} (2020) 124027}
  [\href{https://arxiv.org/abs/2003.01441}{{\ttfamily 2003.01441}}].

\bibitem{Krolak:2004xp}
A.~Krolak, M.~Tinto and M.~Vallisneri, \emph{{Optimal filtering of the LISA
  data}}, \href{https://doi.org/10.1103/PhysRevD.70.022003}{\emph{Phys. Rev. D}
  {\bfseries 70} (2004) 022003}
  [\href{https://arxiv.org/abs/gr-qc/0401108}{{\ttfamily gr-qc/0401108}}].

\bibitem{Vallisneri:2005ji}
M.~Vallisneri, \emph{{Geometric time delay interferometry}},
  \href{https://doi.org/10.1103/PhysRevD.76.109903}{\emph{Phys. Rev. D}
  {\bfseries 72} (2005) 042003}
  [\href{https://arxiv.org/abs/gr-qc/0504145}{{\ttfamily gr-qc/0504145}}].

\bibitem{Wang:2017aqq}
G.~Wang and W.-T. Ni, \emph{{Numerical simulation of time delay interferometry
  for TAIJI and new LISA}},
  \href{https://doi.org/10.1088/1674-4527/19/4/58}{\emph{Res. Astron.
  Astrophys.} {\bfseries 19} (2019) 058}
  [\href{https://arxiv.org/abs/1707.09127}{{\ttfamily 1707.09127}}].

\bibitem{Wang:2020fwa}
G.~Wang and W.-T. Ni, \emph{{Revisiting time delay interferometry for
  unequal-arm LISA and TAIJI}},
  \href{https://doi.org/10.1088/1402-4896/acd882}{\emph{Phys. Scripta}
  {\bfseries 98} (2023) 075005}
  [\href{https://arxiv.org/abs/2008.05812}{{\ttfamily 2008.05812}}].

\bibitem{Wang:2020pkk}
G.~Wang, W.-T. Ni, W.-B. Han and C.-F. Qiao, \emph{{Algorithm for time-delay
  interferometry numerical simulation and sensitivity investigation}},
  \href{https://doi.org/10.1103/PhysRevD.103.122006}{\emph{Phys. Rev. D}
  {\bfseries 103} (2021) 122006}
  [\href{https://arxiv.org/abs/2010.15544}{{\ttfamily 2010.15544}}].

\bibitem{RajeshNayak:2004jzp}
K.~Rajesh~Nayak and J.~Y. Vinet, \emph{{Algebraic approach to time-delay data
  analysis for orbiting LISA}}, .

\bibitem{Nayak:2005un}
K.~R. Nayak and J.~Y. Vinet, \emph{{Algebraic approach to time-delay data
  analysis: Orbiting case}},
  \href{https://doi.org/10.1088/0264-9381/22/10/040}{\emph{Class. Quant. Grav.}
  {\bfseries 22} (2005) S437}.

\bibitem{Tinto:2004wu}
M.~Tinto and S.~V. Dhurandhar, \emph{{TIME DELAY}},
  \href{https://doi.org/10.12942/lrr-2005-4}{\emph{Living Rev. Rel.} {\bfseries
  8} (2005) 4} [\href{https://arxiv.org/abs/gr-qc/0409034}{{\ttfamily
  gr-qc/0409034}}].

\bibitem{Tinto:2014lxa}
M.~Tinto and S.~V. Dhurandhar, \emph{{Time-Delay Interferometry}},
  \href{https://doi.org/10.12942/lrr-2014-6}{\emph{Living Rev. Rel.} {\bfseries
  17} (2014) 6}.

\bibitem{Tinto:2020fcc}
M.~Tinto and S.~V. Dhurandhar, \emph{{Time-delay interferometry}},
  \href{https://doi.org/10.1007/s41114-020-00029-6}{\emph{Living Rev. Rel.}
  {\bfseries 24} (2021) 1}.

\bibitem{Muratore:2020mdf}
M.~Muratore, D.~Vetrugno and S.~Vitale, \emph{{Revisitation of time delay
  interferometry combinations that suppress laser noise in LISA}},
  \href{https://doi.org/10.1088/1361-6382/ab9d5b}{\emph{Class. Quant. Grav.}
  {\bfseries 37} (2020) 185019}
  [\href{https://arxiv.org/abs/2001.11221}{{\ttfamily 2001.11221}}].

\bibitem{Larson:2002xr}
S.~L. Larson, R.~W. Hellings and W.~A. Hiscock, \emph{{Unequal arm space borne
  gravitational wave detectors}},
  \href{https://doi.org/10.1103/PhysRevD.66.062001}{\emph{Phys. Rev. D}
  {\bfseries 66} (2002) 062001}
  [\href{https://arxiv.org/abs/gr-qc/0206081}{{\ttfamily gr-qc/0206081}}].

\bibitem{Cornish:2001qi}
N.~J. Cornish and S.~L. Larson, \emph{{Space missions to detect the cosmic
  gravitational wave background}},
  \href{https://doi.org/10.1088/0264-9381/18/17/308}{\emph{Class. Quant. Grav.}
  {\bfseries 18} (2001) 3473}
  [\href{https://arxiv.org/abs/gr-qc/0103075}{{\ttfamily gr-qc/0103075}}].

\bibitem{Robson:2018ifk}
T.~Robson, N.~J. Cornish and C.~Liu, \emph{{The construction and use of LISA
  sensitivity curves}},
  \href{https://doi.org/10.1088/1361-6382/ab1101}{\emph{Class. Quant. Grav.}
  {\bfseries 36} (2019) 105011}
  [\href{https://arxiv.org/abs/1803.01944}{{\ttfamily 1803.01944}}].

\bibitem{Vallisneri:2007xa}
M.~Vallisneri, J.~Crowder and M.~Tinto, \emph{{Sensitivity and
  parameter-estimation precision for alternate LISA configurations}},
  \href{https://doi.org/10.1088/0264-9381/25/6/065005}{\emph{Class. Quant.
  Grav.} {\bfseries 25} (2008) 065005}
  [\href{https://arxiv.org/abs/0710.4369}{{\ttfamily 0710.4369}}].

\bibitem{Muratore:2022nbh}
M.~Muratore, O.~Hartwig, D.~Vetrugno, S.~Vitale and W.~J. Weber,
  \emph{{Effectiveness of null time-delay interferometry channels as instrument
  noise monitors in LISA}},
  \href{https://doi.org/10.1103/PhysRevD.107.082004}{\emph{Phys. Rev. D}
  {\bfseries 107} (2023) 082004}
  [\href{https://arxiv.org/abs/2207.02138}{{\ttfamily 2207.02138}}].

\bibitem{Nam:2022rqg}
D.~Q. Nam, Y.~Lemiere, A.~Petiteau, J.-B. Bayle, O.~Hartwig, J.~Martino et~al.,
  \emph{{TDI noises transfer functions for LISA}},
  \href{https://arxiv.org/abs/2211.02539}{{\ttfamily 2211.02539}}.

\bibitem{Zhang:2019oet}
C.~Zhang, Q.~Gao, Y.~Gong, D.~Liang, A.~J. Weinstein and C.~Zhang,
  \emph{{Frequency response of time-delay interferometry for space-based
  gravitational wave antenna}},
  \href{https://doi.org/10.1103/PhysRevD.100.064033}{\emph{Phys. Rev. D}
  {\bfseries 100} (2019) 064033}
  [\href{https://arxiv.org/abs/1906.10901}{{\ttfamily 1906.10901}}].

\bibitem{Vecchio:2004ec}
A.~Vecchio and E.~D.~L. Wickham, \emph{{The Effect of the LISA response
  function on observations of monochromatic sources}},
  \href{https://doi.org/10.1103/PhysRevD.70.082002}{\emph{Phys. Rev. D}
  {\bfseries 70} (2004) 082002}
  [\href{https://arxiv.org/abs/gr-qc/0406039}{{\ttfamily gr-qc/0406039}}].

\end{thebibliography}

\providecommand{\href}[2]{#2}\begingroup\raggedright\endgroup

\end{sloppypar}
\end{document}